\documentclass[preprint,12pt]{elsarticle}
\usepackage{amssymb}
\usepackage{hyperref}
\usepackage{amsmath}
\usepackage{xcolor}
\usepackage{url}
\providecommand{\urlprefix}{}
\journal{Physics of the Dark Universe}

\begin{document}

\begin{frontmatter}

\title{Cosmological bounds on dark matter annihilation using dark ages 21-cm signal}

\author{Vivekanand Mohapatra}

%% Author affiliation
\affiliation{organization={Department of Physics},
            addressline={National Institute of Technology Meghalaya}, 
            city={Cherrapunji},
            postcode={793108}, 
            state={Meghalaya},
            country={India}}

%% Abstract
\begin{abstract}
We investigate the impact of dark matter (DM) annihilation on the global 21-cm signal during the dark ages and cosmic dawn eras. The 21-cm line provides a complementary probe for studying the nature of dark matter beyond standard cosmological observables. In the standard $\Lambda$CDM framework, the expected absorption amplitude of the dark ages global 21-cm signal is approximately $-42\, \rm mK$. However, energy injection from DM annihilation can significantly heat and ionize the intergalactic medium, potentially altering or even erasing this absorption feature. We evaluate the thermal and ionization history of the gas to derive an upper bound on $f_\chi^2 \langle \sigma v \rangle / M_\chi$ using the dark ages signal, which is free from astrophysical uncertainties. After incorporating observational and theoretical uncertainties arising from future lunar-based experiments and variations in cosmological parameters, respectively--- we obtain a conservative upper limit of $f_\chi^2\langle\sigma v\rangle/M_\chi \lesssim 10^{-27}~\rm cm^3\,s^{-1}\,\rm GeV^{-1}$. This constraint is stronger than the bounds derived from Planck (2018) data for mass $\lesssim 10~\rm GeV$.
\end{abstract}

%% Keywords
\begin{keyword}
cosmology \sep dark matter \sep dark ages \sep cosmic dawn

\end{keyword}

\end{frontmatter}

\section{Introduction}\label{sec:intro}

The nature of dark matter (DM) remains one of the most profound unresolved problems in particle physics and cosmology. Although its existence is widely supported by various astrophysical and cosmological observations, it remains undetected via non-gravitational means~\cite{Arbey:2021gdg}. One of the earliest and most prominent pieces of evidence for DM comes from the discrepancy between the observed rotation curves of spiral galaxies and the distribution of visible matter~\cite{Persic:1995ru}. Furthermore, in galaxy clusters, a significant amount of hot gas emits X-rays through Bremsstrahlung radiation, influenced by the cluster's gravitational potential. On modeling these clusters and analyzing the X-ray emissions, one can infer the mass of visible matter. However, such analyses consistently reveal the presence of a substantial amount of invisible mass~\cite{Arbey:2021gdg}. Similarly, several measurements of the angular radius of the so-called Einstein ring, producing the distorted image of a distant galaxy on a galaxy cluster, have also hinted towards missing visible mass \cite{Clowe:2006eq}.

On cosmological scales, the $\Lambda$CDM model requires the presence of DM to explain the formation of large-scale structures in the early universe. Simulations of cosmic structure formation, incorporating the interplay between the expansion of the universe and gravity, typically assume DM to be cold and collisionless~\cite{Geller:1989da, 2DFGRS:2001zay, SDSS:2000hjo, Springel:2005nw}. Moreover, DM leaves an imprint on the primordial matter density fluctuations, which can be inferred from the temperature anisotropies in the cosmic microwave background (CMB) radiation~\cite{Planck:2018vyg}. On accounting these signatures, we can state that if DM consists of non-standard particles, it must be long-lived, electrically neutral, massive, and non-relativistic. Alternative scenarios such as warm DM~\cite{Chatterjee:2019jts, Natwariya:2022xlv}, fuzzy DM~\cite{Matos:1998vk, Hu:2000ke, Arbey:2001qi, Arbey:2003sj, Hui:2016ltb}, dark fluid models~\cite{Boyle:2001du, Bilic:2001cg, Guo:2005qy, Arbey:2006it}, and self-interacting DM~\cite{Carlson:1992fn, Spergel:1999mh, Bhatt:2019qbq} have also been proposed, some of which could partially contribute to the total DM density as well. DM candidates such as weakly interacting massive particles (WIMPs) and primordial black holes \cite{Dasgupta:2019cae, Natwariya:2021xki, Mittal:2021egv, Saha:2021pqf}, can annihilate and decay into photons and other standard model particles, resulting in depositing energy into the intergalactic medium (IGM) in the early universe. These exotic energy injections in the sub-GeV range can modify the evolution of the universe, leaving imprints on the CMB anisotropies and power spectrum \cite{Adams:1998nr, Chen:2003gz, Padmanabhan:2005es, Slatyer:2009yq, Kanzaki:2009hf, Slatyer:2015jla, Slatyer:2016qyl, Poulin:2016anj, Cang:2020exa}. In this work, we focus specifically on the effects of DM annihilation on the thermal and ionization evolution of the IGM and its impact on the global 21-cm signal.

In the post-recombination era, the universe was primarily filled with neutral hydrogen atoms and some residual free electrons and protons. Any exotic energy injection can modify the thermal and ionization evolution of the IGM, subsequently altering the hydrogen 21-cm line during the cosmic dawn era \cite{Pritchard:2005an,  Furlanetto:2006fs, Furlanetto:2006jb, Pritchard:2011xb}. Additionally, the structure formation giving birth to the first star during the cosmic dawn can also affect the 21-cm signal, sensitive to the underlying astrophysics of the luminous object \cite{Pritchard:2005an, Furlanetto:2006jb, Pritchard:2011xb}. Recent observation of the global 21-cm signal by the Experiment to Detect the Global Epoch of Reionization Signature (EDGES) \cite{Bowman:2018yin}, reported a tentative absorptional amplitude of $-0.5^{+0.2}_{-0.5}\, \rm K$ at redshift $z\sim 17$, which is twice stronger than the $\Lambda\rm CDM$ predicted signal \cite{Barkana:2018lgd}. However, this detection has remained disputed in the literature, based on modelling of the EDGES data \cite{Hills:2018vyr}, calibration systematic errors \cite{Sims:2019kro}, and possible systematic artifact within the ground plane \cite{Bradley:2018eev}. Moreover, SARAS3 has also rejected the entire signal with a 95.3\% confidence level after conducting an independent check \cite{Singh:2021mxo}.

In addition to the cosmic dawn signal, the standard cosmology also predicts a relatively faint absorption feature in the global 21-cm signal during the dark ages, corresponding to redshifts $z \sim 40$--$250$~($\nu \sim 35$--$5~\rm MHz$). However, this signal remains undetected due to several challenges, including ground-based instrumental limitations, strong and poorly characterized foregrounds, and the Earth's ionosphere, which becomes nearly opaque at frequencies below $\sim 40$ MHz~\cite{Furlanetto:2006jb, Pritchard:2011xb}. Although future ground-based experiments such as the proposed Mapper of the IGM Spin Temperature (MIST) aim to observe the signal in the $25$--$105$ MHz range~\cite{Monsalve:2023lvo}, ionospheric distortion could make it significantly challenging at frequencies below $\sim 40$ MHz. Overcoming these limitations is one of the primary motivations behind several proposed lunar and space-based missions—including FARSIDE~\cite{farside}, DAPPER~\cite{dapper}, FarView~\cite{dapper}, SEAMS~\cite{seams}, and LuSee Night~\cite{luseenight,10906958}—which are designed to operate above the Earth’s ionosphere and far from terrestrial radio frequency interference (RFI), thereby significantly enhancing the prospects for detecting the dark ages 21-cm signal. The recently proposed LuSee Night mission, scheduled for deployment to the farside of the Moon in 2026, aims to observe the sky over a frequency range of $0.1-50\,\rm MHz$, potentially enabling the first detection of the global 21-cm signal from the dark ages~\cite{10906958}. For these future lunar-based experiments, an integration time of 20,000 hours is expected to achieve an uncertainty in the brightness temperature measurement of approximately $\Delta T_b \sim 15\,\rm mK$. Extending the integration time to $10^5$ hours may reduce this uncertainty to $\sim 5\,\rm mK$~\cite{Burns:2020gfh, Rapetti:2019lmf}, making the detection of the standard dark ages signal feasible with high precision.

In this work, we study the effect of DM annihilation on the evolution of IGM temperature and the global 21-cm signal in the post-recombination era. We consider a scenario in which DM annihilates into photons $(\chi\chi \to \gamma\gamma)$ or an electron-positron pair $(e^-e^+)$. We evaluate the global 21-cm signal during both the dark ages and the cosmic dawn for a fiducial star formation history. The $\Lambda$CDM framework predicts the absence of luminous astrophysical sources during the dark ages, making this epoch an ideal probe for exotic physics, free from astrophysical uncertainties. In contrast, the 21-cm signal during the cosmic dawn is strongly influenced by the properties of early astrophysical sources; we demonstrate that the inclusion/exclusion of an excess radio background leads to a wide range of possible signal shapes and depths. Our results show that, even in this minimalistic setup, DM annihilation can inject sufficient energy at high redshifts to modify or even erase the dark ages signal. After accounting for observational uncertainties from upcoming lunar-based experiments and theoretical uncertainties arising from variations in cosmological parameters, we derive upper bounds on $f_\chi^2 \langle\sigma v\rangle / M_\chi$ that are free from astrophysical uncertainties and stronger than the limits obtained from Planck (2018) data \cite{Planck:2018vyg} for mass $\lesssim 10~\rm GeV$.

The paper is organised as follows. In Sec.~\ref{sec: G21cm}, we discuss the evolution of global 21-cm signal from the dark ages and the cosmic dawn era, for a standard star formation history. In Sec.~\ref{sec: Tgas}, we study the thermal and ionization evolution of IGM in the presence of DM annihilation. Further, in Sec.~\ref{sec: Result}, we investigate the effect of DM annihilation on the dark ages global 21-cm signal, and derive upper bounds. Finally, in Sec.~\ref{sec: conclusion}, we conclude our work with the existing bounds in the literature, and discuss a future outlook.

\section{Global 21-cm signal}\label{sec: G21cm}
The hyperfine interaction in neutral hydrogen (HI) atoms splits the ground state into singlet $(\rm F = 0)$ and triplet $(\rm F = 1)$ states. The relative population density of HI in triplet $(n_1)$ and singlet $(n_0)$ states is given by ${n_1}/{n_0} = {g_1}/{g_0} \,\exp\{-T_*/T_s\}$, where $g_0 = 1$ and $g_1 = 3$ represent the statistical weights of the singlet and triplet states, respectively. $T_* = 68\,\rm mK$ is the equivalent temperature of the photons produced from the hyperfine transition. These photons have a wavelength of 21 cm (or frequency $\nu_{21} = 1420\,\rm MHz$). The quantity $T_s$, known as the spin temperature, characterizes the relative population distribution between these two levels.

The intensity distribution of 21-cm photons in the CMB can be approximated by the Rayleigh-Jeans law. The equivalent temperature of the redshifted global (sky-averaged) intensity relative to the CMB can be expressed as $T_{21} = [(T_s-T_{R})/1+z]\left(1-e^{-\tau_{21}}\right)$, where $T_R$ represents the background radiation. $\tau_{21}$ represents the optical depth of the 21-cm photons. In $\Lambda\rm CDM$ framework, $T_R = T_\gamma$; where $T_{\gamma} = T_{\gamma,0}(1+z)$ is the CMB temperature with the present-day value $T_{\gamma,0} = 2.725\,\rm K$. In Sec. \ref{Sec:CD}, we will discuss a possible scenario that can produce an excess-radio background, resulting in $T_R>T_\gamma$. Now, in the limit of $\tau_{21}\ll 1$, the above equation can be expressed as \cite{Furlanetto:2006fs,Mesinger:2007pd, Mesinger:2010ne,  Pritchard:2011xb}

\begin{equation}
    T_{21} \approx 27~\mathrm{mK}~ x_{\rm HI} \left(\frac{\Omega_{b,0}h^2}{0.023}\right) \Bigg(\frac{0.14}{\Omega_{m,0}h^2} \frac{1+z}{10}\Bigg)^{1/2} \left(1-\frac{T_{R}}{T_s}\right)\,
     \label{eq:T21}
\end{equation}

Here, $\Omega_{b,0}$ and $\Omega_{m,0}$ are the present-day baryon and total matter density parameters, respectively, whereas $h = 0.67$ represents the Hubble parameter in the unit $100\,\rm Km/s/Mpc$ \cite{Planck:2018vyg}. $x_{\rm HI}$ is the neutral hydrogen fraction which can approximated as $\sim(1-x_e)$, where $x_e$ represents the ionization fraction. The post-recombination era $(z\lesssim 1100)$ primarily consists of HI atoms and some residual free electrons and protons; thus, determining the evolution of $T_{21}$ can provide pristine cosmological information, and the astrophysical nature of the first star \cite{Pritchard:2011xb}. From the above equation, we note that $T_{21}$ primarily depends on $x_e$ and the ratio $(T_{R}/T_s)$. The evolution of the spin temperature is given by \cite{Furlanetto:2006fs,Mesinger:2007pd, Mesinger:2010ne,  Pritchard:2011xb}

\begin{equation}
	T_s^{-1} = \frac{T_{R}^{-1}+x_{\alpha}T_{\alpha}^{-1}+x_cT_g^{-1}}{1+x_{\alpha}+x_c}\, ,
	\label{eq:spin temp}
\end{equation}
where, $T_g$ and $T_{\alpha}$ represent the gas and color temperatures, respectively. Usually, the color temperature $T_{\alpha}\approx T_g$, because the optical depth of Ly$\alpha$ photons is large, which leads to a large number of scatterings, thus bringing the radiation field and IGM to local thermal equilibrium \cite{Furlanetto:2006fs, Pritchard:2005an}. $x_c$ and $x_{\alpha}$ represent the collisional and $\rm Ly\alpha$ couplings, respectively \cite{1952AJ.....57R..31W, 1959ApJ...129..536F, 1958PIRE...46..240F}. From Eq. \eqref{eq:T21}, we can observe that for $T_s<T_{R}$ one expects an absorptional signal. In the sections below, we will explain the two absorptional signals predicted in the $\Lambda\rm CDM$ framework.

\subsection{Dark ages signal}

In the post-recombination era $(z\lesssim 1100)$, the gas was coupled to the CMB via efficient inverse Compton scattering between the electrons/protons and the CMB photons. Consequently, the gas and CMB shared the same temperature, resulting in an absence of $T_{21}$ signal. However, at redshifts $z\lesssim 250$ the Compton scattering became ineffective, resulting in the evolution of $T_g$ and $T_R$ as $(1+z)^2$ and $(1+z)$, respectively, due to adiabatic expansion. Meanwhile, efficient collisional coupling between the electrons/protons and HI atoms kept the spin temperature coupled to gas temperature till $z\sim 40$--- resulting in $T_s<T_R$. We expect an absorption signal at redshifts $z\sim 250-40$, referred to as the dark ages global 21-cm signal \cite{Furlanetto:2006fs, Pritchard:2011xb}. The collisional coupling coefficient $(x_c)$ is given by \cite{2006nla..conf..296Z, 2007MNRAS.374..547F, 2007MNRAS.379..130F},

\begin{equation*}
    x_c = \frac{T_*}{T_R}\,\frac{n_ik^{i\rm H}_{10}}{A_{10}},
\end{equation*}
where $A_{10} = 2.85 \times 10^{-15}\,\rm Hz$ is the Einstein coefficient for spontaneous emission in the hyperfine state. $n_i$ represents the number density of the species ``$i$" present in the IGM while $k^{i\rm H}_{10}$ represents their corresponding collisional spin deexcitation rate. The deexcitation rates $k^{eH}_{10}$ and $k^{HH}_{10}$ can be approximated in a functional form as follows \cite{2006ApJ...637L...1K, 2001A&A...371..698L, Mittal:2021egv, Pritchard:2011xb}

\begin{alignat}{2}
	k^{HH}_{10} & = 3.1 \times 10^{-17}\left(\frac{T_{g}}{\mathrm{K}}\right)^{0.357}\cdot e^{-32\mathrm{K} / T_{g}}, \\
	\log_{10}{k^{eH}_{10}} & = -15.607 + \frac{1}{2}\log_{10}\left(\frac{T_{g}}{\mathrm{K}}\right)\times \exp\left\{-\dfrac{\left[\log_{10} \left(T_{g}/\mathrm{K}\right)\right]^{4.5}}{1800}\right\}\, .
\end{alignat}

Here, all the $k^{iH}_{10}$ terms are in the dimension of $\rm m^3s^{-1}$ and have been approximated for $T_g\lesssim 10^4\,\rm K$. Further, at redshifts $z\lesssim 40$, $x_c$ becomes ineffective causing $T_s$ approach $T_R$--- resulting in $T_{21}\simeq 0\,\rm mK$. Afterwards, the formation of the first star emits copious Ly$\alpha$ and X-ray radiation that leads to $T_{21}<0$ at redshifts $z\lesssim 30$. In the next section, we discuss the effect of Ly$\alpha$ radiation on the evolution of 21-cm signal.

\subsection{Cosmic dawn signal}\label{Sec:CD}

After star formation commenced, the radiation from the first luminous sources began to heat and ionize the intergalactic medium. In particular, Ly$\alpha$ photons can induce hyperfine transitions in neutral hydrogen atoms through the Wouthuysen–Field effect \cite{1952AJ.....57R..31W, 1959ApJ...129..536F}, effectively coupling the spin temperature to the gas temperature. At $T_s<T_R$, we expect an absorption signal at redshifts $z\lesssim 30$, till the universe becomes ionized again \cite{Furlanetto:2006jb, Pritchard:2011xb}.

The Ly$\alpha$ coupling coefficient, which quantifies the strength of the Wouthuysen–Field effect, can be expressed as \cite{Hirata:2005mz, Mesinger:2010ne, Pritchard:2011xb}
\begin{equation}
    x_{\alpha} = \frac{T_{*}}{T_R}\frac{4P_{\alpha}}{27 A_{10}},
\end{equation}
where $P_{\alpha}$ is the total scattering rate of Ly$\alpha$ photons per neutral hydrogen atom. This rate depends on the specific intensity $J_{\alpha}$ of the Ly$\alpha$ radiation field, which in turn requires knowledge of the Ly$\alpha$ photon emissivity, $\epsilon_{\alpha}$ \cite{Barkana:2004vb}. Assuming the stellar sources are predominantly Pop II stars, we adopt their spectral energy distribution (SED) as $\phi(\alpha) = 2902.91\, \Tilde{E}^{-0.86}$ \cite{Mittal:2020kjs}, where $\Tilde{E} = E/E_{\rm ion}$ and the photon energy $E$ lies within $[E_{\alpha}, E_{\beta}]$, with $E_{\alpha} = 10.2\,\rm eV$, $E_{\beta} = 12.09\,\rm eV$, and $E_{\rm ion} = 13.6\,\rm eV$ denoting the energies corresponding to the Ly$\alpha$, Ly$\beta$, and Lyman-limit transitions, respectively. The Ly$\alpha$ emissivity is then given by \cite{Mittal:2020kjs}
\begin{equation}
    \epsilon_{\alpha}(E, z) = \frac{f_{\alpha} \phi_{\alpha}(E)\, \dot{\rho}_{*}(z)}{m_b},
    \label{eq:Ly_alpha_emissivity}
\end{equation}
where $m_b$ is the baryon mass, $f_{\alpha}$ is a scaling factor for the spectrum $\phi_{\alpha}$, and $\dot{\rho}_{*}$ denotes the star formation rate density (SFRD), which is governed by the rate of baryon collapse into dark matter haloes \cite{Barkana:2004vb}. The SFRD can be modelled as
\begin{equation*}
    \dot{\rho}_{*}(z) = -f_{*} \bar{\rho}_{b}^{0}(1 + z)H(z) \frac{dF_{\rm coll}(z)}{dz}\, ,
\end{equation*}
where $\bar{\rho}_{b}^{0} = \rho_c \Omega_{b,0}$ is the present-day baryon density, and $f_{*}$ represents the star formation efficiency. The collapsed baryon fraction $F_{\rm coll}(z)$ is obtained using the Press–Schechter formalism \cite{Press:1973iz, Barkana:2000fd}
\begin{equation}
    F_{\rm coll}(z) = \mathrm{erfc}\left[\frac{\delta_c(z)}{\sqrt{2}\sigma(m_{\rm min})}\right],
    \label{eq:F_coll}
\end{equation}
where $\delta_c$ is the linear critical overdensity for collapse, $\sigma^2$ is the variance of the smoothed density field, and $\mathrm{erfc}(\cdot)$ is the complementary error function. The minimum halo mass, $m_{\rm min}$, associated with a given virial temperature $\rm T_{\rm vir}$, determines the smallest haloes capable of hosting star formation. This relation is given by \cite{Mittal:2021egv}
\begin{equation}
    m_{\rm min} = \frac{10^8\, \rm M_{\odot}}{\sqrt{\Omega_m h^2}}\left[\frac{10}{1+z}\frac{0.6}{\mu} \frac{\rm min(T_{\rm vir})}{1.98\times 10^4\,\rm K}\right]^{3/2},
\end{equation}
where $\rm M_{\odot}$ is the solar mass, and $\mu \approx 1.22$ is the mean molecular weight \cite{DAYAL20181}. In this work, we consider haloes with $\rm T_{\rm vir} \geq 10^4\,\rm K$ as the threshold for star formation. The parameter $\delta_c/\sigma(m_{\rm min})$ is evaluated using the \texttt{COLOSSUS} software \cite{Diemer:2017bwl}. Once the SFRD is defined, we can compute the Ly$\alpha$ specific intensity $J_{\alpha}$ as \cite{Hirata:2005mz, Pritchard:2011xb}
\begin{equation}
    J_{\alpha} = \frac{c}{4\pi}(1+z)^2 \sum_{n = 2}^{23} P_n \int_{z}^{z_{\rm max}} \frac{\epsilon_{\alpha}(E_n', z')}{H(z')} \, dz,
    \label{eq:J_alpha}
\end{equation}
where $P_n$ is the probability that a photon from the $n^{\rm th}$ Lyman series transition redshifts to the Ly$\alpha$ frequency before being absorbed or scattered. The values of $P_n$ are tabulated in the article \cite{Hirata:2005mz, Pritchard:2005an}. The energy of a photon emitted at redshift $z'$ and observed at redshift $z$ is redshifted as $E_n' = E_n (1+z')/(1+z)$, where $E_n$ corresponds to the energy of a transition from the $n^{\rm th}$ level to the ground state of hydrogen. The upper limit of the redshift integral in Eq.~\eqref{eq:J_alpha} is determined by the maximum redshift from which a photon can redshift into the Ly$\alpha$ line
\begin{equation}
    1 + z_{\rm max} = \frac{E_{n+1}}{E_n}(1 + z) = \frac{1 - (1 + n)^{-2}}{1 - n^{-2}}(1 + z).
\end{equation}

We then rewrite the Ly$\alpha$ coupling coefficient as \cite{Mittal:2020kjs}
\begin{equation}
    x_{\alpha} = \frac{S J_{\alpha}}{J_0},
\end{equation}
where $J_0 \approx 5.54\times 10^{-8}\, (T_R/T_{\gamma})\,\rm m^{-2}\,s^{-1}\,Hz^{-1}\,sr^{-1}$, and $S$ is the scattering correction factor. In this work, we assume $S \sim 1$.

The tentative twice-than-expected amplitude of global 21-cm signal reported by the EDGES collaboration hints towards the presence of excess radio background in the early universe \cite{Bowman:2018yin}. Additionally, the ARCADE2 and LWA1 observations also reported a power-law-like radio background, supporting the possibility of the presence of radio sources in the early universe \cite{Fixsen:2009xn, Feng:2018rje, Dowell:2018mdb}. Although the EDGES observation is disputed and has been challenged in the literature \cite{Hills:2018vyr, Bradley:2018eev, Sims:2019kro, Bevins:2022clu}, the possibility of an excess radio background is not completely ruled out.

We model the ERB assuming that the early galaxy formation produces radio radiation proportional to the star formation rate (SFR). In particular, we considered the radio luminosity $(L_R)$ and SFR $(\dot{M}_*)$ relation at 150 MHz that can hold at higher redshift as \cite{10.1093/mnras/sty016}

\begin{equation}
    L_R = 10^{22} \zeta_{R}\left(\frac{\dot{M}_*}{1~M_{\odot}~\rm yr^{-1}}\right)~\rm J\,s^{-1}\, Hz^{-1}\, ,
\end{equation}
where $ \zeta_{R}$ is a free parameter that accounts for the difference between the local observation and the observation of the high-redshift universe. The relation can be extrapolated further, assuming a spectral index of $-0.7$ \cite{10.1093/mnras/sty016}. The globally averaged radio luminosity at 150 MHz per unit comoving
volume at redshift $z$ is then given by

\begin{equation}
    \epsilon_{R, 150\,\rm MHz}(z) = 10^{22}\zeta_Rf_*\bar{\rho}_m\frac{dF_{\rm coll}}{dt}\, \rm J\,s^{-1}Mpc^{-3}Hz^{-1}
\end{equation}
where $\bar{\rho}_m$ represents the average baryon density. Now, the 21-cm radiation flux can be expressed as \cite{Ciardi:2003hg, Chatterjee:2019jts,  Choudhury:2020azd}

\begin{equation}
    F_R (z) = 10^{22} \left(\frac{1420}{150}\right)^{-0.7} \frac{(1+z)^{3.7}}{4\pi}\zeta_R f_*\bar{\rho}_m \times\int_z^{\infty}\frac{1}{(1+z')^{0.7}}\frac{dF_{\rm coll}(z')}{dz'}\, dz'\,.
     \label{eq:radio_flux}
\end{equation}
We then convert the flux into brightness temperature, $T_{\rm ERB}$ and add it to the CMB to obtain $T_{R} = T_{\gamma}+ T_{\rm ERB}$.

\section{Evolution of IGM in the presence of DM annihilation}\label{sec: Tgas}

The thermal evolution of the universe in the presence of energy injection from DM annihilation can be expressed as \cite{Peebles:1968ja, Seager:1999bc, Seager:1999km, DAmico:2018sxd, Mitridate:2018iag, Short:2019twc, PhysRevD.110.123506}

\begin{equation}
    \frac{dT_g}{dz} = 2\frac{T_g}{1+z}  + \frac{\Gamma_c}{(1+z)H(z)} (T_g - T_\gamma)\,
     + \frac{2}{3n_bk_B(1+z)H(z)}[ Q_{\rm X}+Q_{\chi}]\, ,
    \label{Gas_Evolution}
\end{equation}
where $H$ and $k_B$ represent the Hubble parameter and Boltzmann constant, respectively. $n_{b} = n_H(1+f_{He}+x_e)$ represents the total number density of gas--- here $n_H$, $f_{He} = 0.08$, and $x_e$ are the hydrogen number density, helium fraction, and ionization fraction, respectively \cite{PhysRevD.110.123506, Mohapatra:2023ury}. The first and second terms in the above equation account for the adiabatic cooling of the gas and coupling between the CMB and gas via inverse Compton scattering. The third term contributes to the heating of the gas from the X-ray $(Q_{\rm X})$ radiation and dark matter annihilation $(Q_{\chi})$. Here, $Q_{\rm i}$ represents the energy density rate of the energy sources. The Compton scattering rate $(\Gamma_c)$ is given by

\begin{equation*}  
    \Gamma_c = \frac{8 x_e\sigma_T a_rT_{\gamma}^4 (z)}{3m_e (1+f_{He}+x_e)}\, ,
    \label{Compton_scattering}
\end{equation*}
where $m_e$ and $\sigma_T$ represent the rest mass of an electron and Thomson scattering cross-section, respectively. Whereas, $a_r = 7.57\times 10^{-16}$~J\,$\text{m}^{-3}\,\text{K}^{-4}$ represents the radiation density constant. The evolution of the ionization fraction can be expressed as \cite{Peebles:1968ja, Seager:1999bc, PhysRevD.110.123506, Mohapatra:2023ury}

\begin{equation}
	\frac{dx_e}{dz} = \frac{\mathcal{P}}{(1+z)H} \left[n_Hx_e^2\alpha_B - (1-x_e)\beta_Be^{-E_{\alpha}/k_BT_{\gamma}}\right],
	\label{xe_evolution}
\end{equation}

where $\mathcal{P}$ represents Peebles coefficient, while $\alpha_B$ and $\beta_B$ are the case-B recombination and photo-ionization rates, respectively \citep{Seager:1999bc, Seager:1999km, Mitridate:2018iag}. The Peebles coefficient is given by \cite{Peebles:1968ja, DAmico:2018sxd}
\begin{equation*}
    \mathcal{P} = \frac{1+ \mathcal{K}_H\Lambda_Hn_H(1-x_e)}{1+ \mathcal{K}_H(\Lambda_H+\beta_H)n_H(1-x_e)},
\label{peeble_coefficient}
\end{equation*}
where $\mathcal{K}_H = \pi^2/(E_{\alpha}^3H)$, $E_\alpha = 10.2\,\rm eV$, and $\Lambda_H = 8.22\,\rm{sec}^{-1}$ represents redshifting Ly${\alpha}$ photons, rest frame energy of Ly$\alpha$ photon, and 2S-1S level two-photon decay rate in hydrogen atom respectively \cite{PhysRevA.30.1175}. 

Now, let us discuss the role of X-ray photons in the evolution of the gas temperature and ionization fraction. Unlike Ly$\alpha$ photons, X-ray photons have a much longer mean free path, allowing them to propagate far from their sources and effectively heat and partially ionize the gas \cite{Mirabel:2011rx}. The primary astrophysical sources of X-ray photons include X-ray binaries and mini-quasars \cite{Madau:2003um, Power:2012hm, Fragos:2013bfa}. The mechanism of X-ray heating proceeds as follows: X-ray photons travel through the gas and photoionize neutral hydrogen atoms, releasing energetic free electrons. These electrons then transfer energy through excitations and collisions with other atoms and residual free electrons, thereby increasing the average kinetic energy of the gas and raising its temperature. To relate the X-ray emissivity with the SFR, we assume that the SFR is proportional to the rate at which baryonic matter collapses into virialized haloes, i.e., $dF_{\rm coll}/dt$ (see Eq.~\ref{eq:F_coll}). Following Ref.~\cite{Furlanetto:2006jb}, we express the X-ray heating term in Eq.~\eqref{Gas_Evolution} as

\begin{equation}
    \frac{2}{3}\frac{Q_{\rm X}}{k_Bn_b(1+z)H(z)} = 5\times 10^5\,\mathrm{K}\,(f_Xf_*f_{Xh}) \frac{dF_{\rm coll}}{dz},
    \label{X-ray_term_in_Tg}
\end{equation}

where $f_X$ is a normalization parameter (analogous to $f_{\alpha}$ for Ly$\alpha$ coupling), and $f_{Xh}$ is the fraction of X-ray energy deposited into heating the gas. As $f_X$ and $f_{Xh}$ are degenerate parameters, we treat their product $f_X f_{Xh}$ as a single effective quantity. Next, we investigate the effect of X-ray radiation on the ionization fraction evolution. Since ionizing photons are generated within galaxies, their production rate is considered to scale with the star formation rate \cite{Furlanetto:2006jb}. The ionization efficiency $\xi_{\rm ion}$ can be expressed as

\begin{equation*}
    \xi_{\rm ion} = \mathrm{A_{He}}\, f_* f_{\rm esc} N_{\rm ion},
\end{equation*}

where $f_{\rm esc}$ denotes the fraction of ionizing photons that escape their host galaxies, $N_{\rm ion}$ is the number of ionizing photons produced per baryon, and $\mathrm{A_{He}} = 4/(4 - 3Y_p)$ accounts for the helium mass fraction. The evolution of the ionization fraction in the presence of X-rays is given by \cite{Furlanetto:2006jb, Mohapatra:2023ury}

\begin{equation}
    \frac{dx_e}{dz} = \frac{dx_e}{dz}\bigg|_{\text{Eq.~\eqref{xe_evolution}}} - \xi_{\rm ion} \frac{dF_{\rm coll}}{dz}.
    \label{xe_evolution_modified}
\end{equation}

Since $f_{\rm esc}$ and $N_{\rm ion}$ are degenerate, we treat their product $(f_{\rm esc} N_{\rm ion})$ as a single parameter and set it to unity for simplicity.

Further, we discuss the effect of energy injection from DM annihilation into the IGM. DM can annihilate via different channels such as photons, electron/positron, and other baryons; however, in this work, we consider DM annihilating to produce a pair of photons $(\chi\chi\to\gamma\gamma)$, for simplicity. Now, the DM can deposit energy into the IGM in two ways. First, the energy injection near the decoupling of gas from the CMB can increase the ionization fraction, resulting in a slower rate of hydrogen atom formation. In the $\Lambda\rm CDM$ framework, the ionization fraction at redshift $z\sim 600$ takes a value of $x_e\sim 10^{-3}$ \cite{10.1111/j.1365-2966.2010.16940.x, 10.1111/j.1365-2966.2010.17940.x}; therefore, a larger $x_e$ will delay the decoupling, resulting in a larger $T_g$ at lower redshift. Furthermore, the photons produced from DM annihilation can directly heat the hydrogen atom, resulting in an increase in the average kinetic temperature of the gas. The DM heating term in Eq. \eqref{Gas_Evolution} is given by \cite{Liu:2016cnk, DAmico:2018sxd, Liu:2018uzy, Short:2019twc}

\begin{equation}
    \frac{2}{3}\frac{Q_{\chi}}{k_Bn_b(1+z)H(z)} =  -\frac{2f_{\rm heat}}{3k_Bn_b(1+z)H(z)}\frac{dE}{dVdt}\, .
    \label{DM_term_in_Tg}
\end{equation}
Similarly, the ionization fraction evolution in the presence of DM annihilation is given by \cite{Liu:2016cnk, DAmico:2018sxd, Liu:2018uzy, Short:2019twc}

\begin{equation}
    \frac{dx_e}{dz} = \frac{dx_e}{dz}\bigg|_{\text{Eq.~\eqref{xe_evolution_modified}}} -  \frac{1}{n_H(1+z)H(z)} \frac{dE}{dVdt}\times \Bigg[\frac{\mathcal{P} f_{\rm ion}}{E_0}+\frac{(1-\mathcal{P})f_{\rm exc}}{E_\alpha}\Bigg]\, .
    \label{xe_evolution_DM}
\end{equation}

Here, $E_\alpha = 10.2\,\rm eV$ and $E_0 = 13.6\,\rm eV$ are the Ly$\alpha$ energy and binding energy of the hydrogen atom in its ground state, respectively. The $f_{\rm heat}$, $f_{\rm exc}$, and $f_{\rm ion}$ represent the fraction of energy produced from the DM annihilation that goes to heat, excite, and ionize the IGM. For simplicity, one can consider the instantaneous deposition approximation, i.e., energy produced at any redshift $z$ gets deposited into the IGM immediately. However, in practice, the deposition of energy depends on the redshift, ionization fraction, the annihilation channel, and the energy of the particle. Under the ``SSCK'' approximation proposed in Ref. \cite{Shull:1982zz, Chen:2003gz}, the three terms can be expressed as 

\begin{equation*}
    f_{\rm heat} = f_{\rm eff}\frac{1+2x_e}{3},\qquad\qquad f_{{red}\rm exc} = f_{\rm ion} = f_{\rm eff}\frac{1-x_e}{3}\, ,
\end{equation*}
where, $f_{\rm eff}$ represents the fraction of injected energy deposited via the three processes.
Now, the energy density rate deposited from the DM annihilation can be written as \cite{Liu:2016cnk, DAmico:2018sxd, Liu:2018uzy, Short:2019twc}

\begin{equation}
    \frac{dE}{dVdt} = \rho_{\chi}^2f_{\chi}^2 ~f_c(z, x_e)\frac{\langle\sigma v\rangle}{M_{\chi}}\, ,
    \label{eq:DM_energy_density}
\end{equation}
where $M_{\chi}$, $\langle\sigma v\rangle$, and $f_{\chi}$ represent the mass, annihilation cross-section, and the fraction of DM annihilating. Whereas, $\rho_{\chi} = \Omega_{\chi, 0}\,\rho_c (1+z)^3$ is the dark matter density, with $\Omega_{\chi, 0}$ being the present-day dark matter density parameter. 
The term $f_c(z, x_e)$ encapsulates the fraction of injected energy deposited via the above three processes. In this work, we use the \texttt{PPPC} module of \texttt{DarkHistory}\footnote{\url{https://github.com/hongwanliu/DarkHistory/}} code to calculate $f_c(z, x_e)$ for $\chi\chi\to\gamma\gamma$ and $\chi\chi\to e^{-}e^{+}$ annihilation channels ~\cite{Liu:2019bbm}.
In the next section, we discuss the effect of DM annihilation on the dark ages and cosmic dawn global 21-cm signals. 

\section{Result}
\label{sec: Result}

In this section, we evaluate the effect of DM annihilation on the thermal and ionization evolution of the gas, and subsequently we calculate the global 21-cm signal to obtain upper bounds on the mass and cross-section of dark matter, $\langle\sigma v\rangle-M_{\rm DM}$. To begin with, we first solve Eqs. \eqref{Gas_Evolution} and \eqref{xe_evolution}, in the absence of X-ray and DM heating, simultaneously with initial conditions $T_g = 2758\,\rm K$ and $x_e = 0.05725$ at $z = 1010$ adopted from \texttt{Recfast++} \cite{10.1111/j.1365-2966.2010.16940.x, 10.1111/j.1365-2966.2010.17940.x}. To solve the coupled differential equations, we use \texttt{scipy.integrate.solve\_ivp} solver offered by the \texttt{SciPy}\footnote{\url{https://docs.scipy.org/doc/scipy/reference/index.html}} library in \texttt{Python} programming language. We set the step size at $5000$ and specifically used \texttt{Radau} method, which employs an implicit \texttt{Runge}\texttt{-Kutta} scheme based on Radau IIA quadrature with $5^{\rm th}$-order accuracy, efficient to handle stiff differential equations. 

\begin{figure}[t]
    \centering
        {\includegraphics[width=0.48\textwidth]{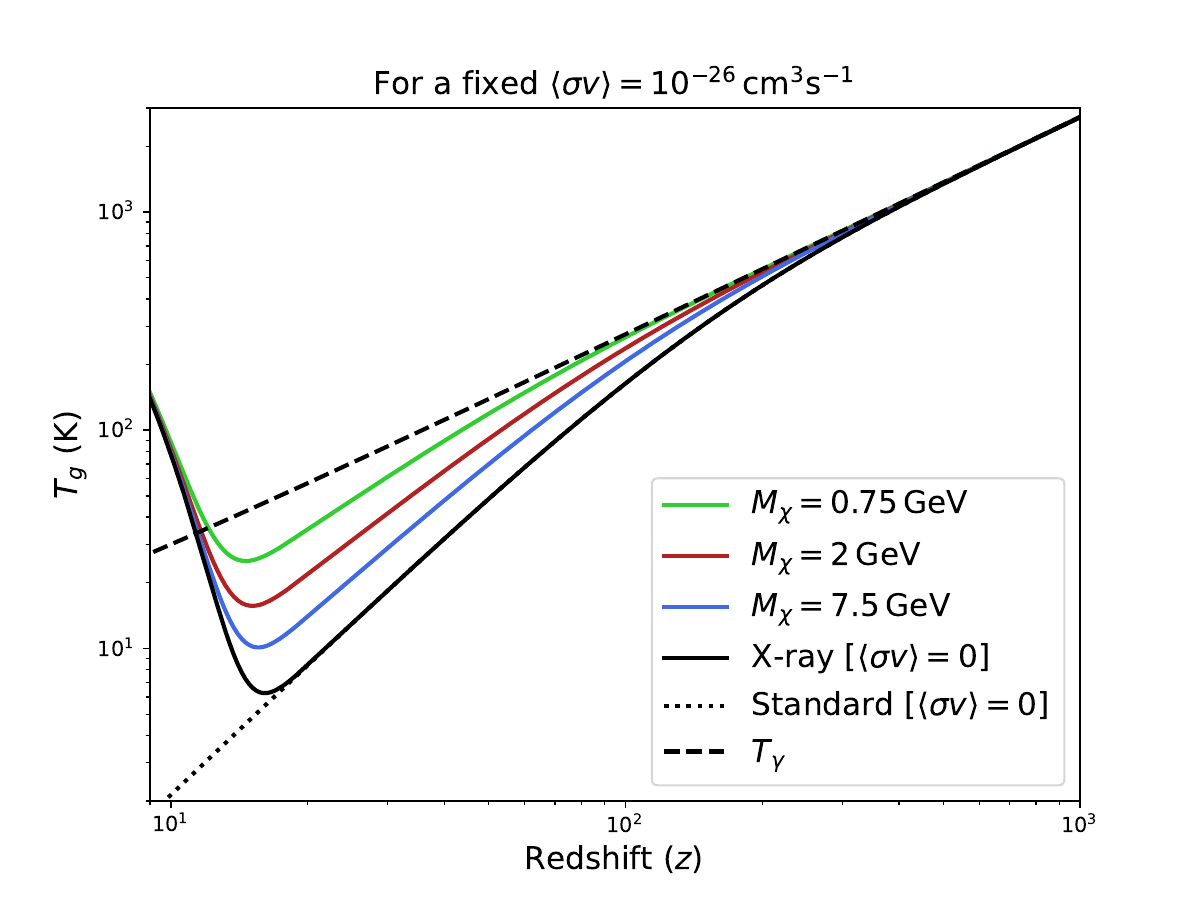}%
        \label{Fig:T_gas}}
    \hfill
        {\includegraphics[width=0.48\textwidth]{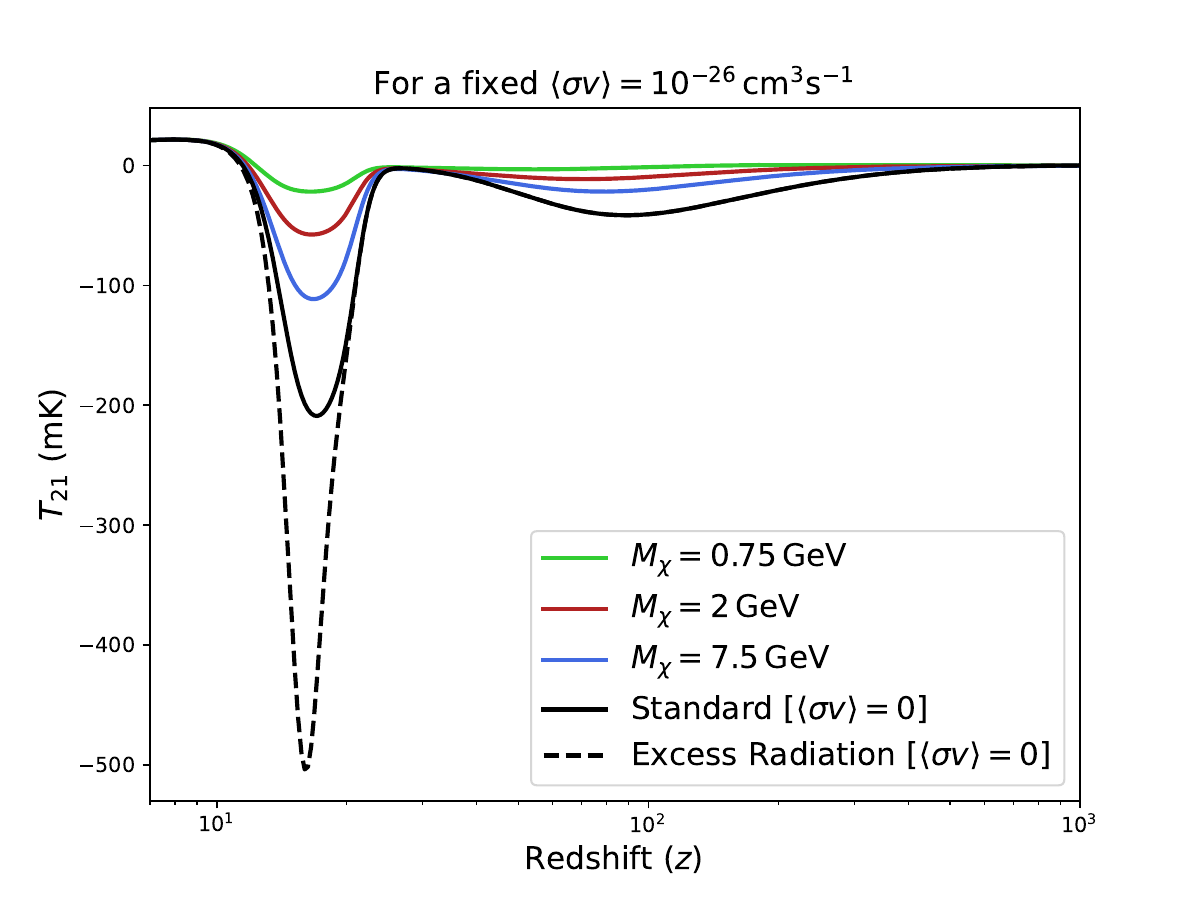}%
        \label{Fig:T21}}
    \caption{Evolution of (a) gas temperature and (b) global 21-cm signal in the presence of DM annihilation after recombination. 
    In panel (a), the black dashed and dotted lines represent the evolution of CMB and gas temperatures in the $\Lambda$CDM framework, respectively. 
    The black solid line shows the IGM temperature evolution in the presence of X-ray heating during the cosmic dawn. 
    The blue, red, and green solid lines correspond to DM annihilation with $M_{\chi} = 7.5$ GeV, $2$ GeV, and $0.75$ GeV, respectively, at a fixed $\langle\sigma v \rangle = 10^{-26}\,\mathrm{cm^3\,s^{-1}}$. 
    The corresponding impact on the $T_{21}$ signal is shown in panel (b) with the same color coding, except for the black dashed line, which illustrates the presence of an excess radio background [see Eq.~\eqref{eq:radio_flux}] without DM annihilation.}
    \label{Fig:DM}
\end{figure}

In Fig.~\ref{Fig:DM}a, we illustrate the evolution of the gas temperature, $T_g$, in the post-recombination era. The black dashed and dotted lines represent the evolution of the CMB temperature and $T_g$ in the standard $\Lambda$CDM scenario. 
We then incorporate X-ray heating into the thermal history, which significantly raises $T_g$ at redshifts $z \lesssim 20$, shown in the solid black line. As described in Eq.~\eqref{X-ray_term_in_Tg}, the magnitude of X-ray heating primarily depends on the parameters $f_*$, $f_X f_{Xh}$, and the virial temperature $T_{\rm vir}$. These quantities depend on the star formation history and the spectral energy distribution of the ionizing source. Therefore, the thermal evolution of the IGM during the cosmic dawn era is largely determined by the underlying astrophysical model. For our fiducial analysis, we adopt $f_* = 0.2$, $f_X f_{Xh} = 0.1$, and $T_{\rm vir} = 7 \times 10^4\,\rm K$. Later in this section, we show that these fiducial values are sufficient to produce the required depth of the 21-cm absorption signal, $T_{21}$.

We further consider the effect of dark matter annihilation on gas heating, as described in Eq.~\eqref{DM_term_in_Tg}. For illustration, we fix the annihilation cross section to $\langle\sigma v\rangle = 10^{-26}\,\rm cm^3\,s^{-1}$ and vary the mass of dark matter. The solid blue, red, and green curves correspond to $M_\chi = 7.5$ GeV, $2$ GeV, and $0.75$ GeV, respectively. We observe that for smaller DM masses, the resulting heating is sufficient to raise $T_g$ to near $T_\gamma$ at redshifts $z \gtrsim 60$, and even exceed it for slightly higher annihilation cross-sections. This trait can be analyzed from Eq.~\eqref{eq:DM_energy_density}, which shows that the energy injection rate $(dE/dVdt)$ scales as $(1+z)^6 \langle\sigma v\rangle/M_\chi$. A lower $M_\chi$ implies a higher DM number density, leading to enhanced annihilation rates and increased photon production at high redshift. Consequently, energy deposition into the gas is more efficient at early times. We also find that $T_g$ rises during the cosmic dawn era due to DM heating, although at a slower rate compared to the dark ages. To further quantify the impact of dark matter annihilation on the thermal and ionization history of the IGM, we compute the global 21-cm signal from both the dark ages and the cosmic dawn.

In Fig.~\ref{Fig:DM}b, we show the evolution of $T_{21}$ as a function of redshift. The black solid curve represents the standard prediction in the $\Lambda$CDM framework. At redshifts $z \gtrsim 250$, no 21-cm signal is observed due to the tight coupling between the gas and the CMB. At later times, around $z \sim 89$, collisional coupling becomes effective, producing a shallow absorption feature with a depth of approximately $-42\,\rm mK$. A more prominent absorption trough appears at $z \sim 17$ with an amplitude of $\sim 208\,\rm mK$, due to the onset of the Wouthuysen–Field effect and X-ray heating. This absorption depth has often been considered to be a standard value in the literature \cite{Barkana:2018lgd}. We obtain this fiducial profile by assuming $f_\alpha = 10$ in Eq.~\eqref{eq:Ly_alpha_emissivity}, which determines the Ly$\alpha$ photon emissivity and hence the coupling of the spin temperature to the gas temperature. Since the properties of the 21-cm signal are sensitive to the star formation history, varying the free parameters associated with Ly$\alpha$ coupling ($f_\alpha$) and X-ray heating ($f_X f_{Xh}$) can lead to different absorption depths, shapes, and peak redshifts. We also include the case where an excess radio background is present during the cosmic dawn, shown as the black dashed line. Using Eq.~\eqref{eq:radio_flux}, we compute the enhanced radio photon flux for a fiducial value of $\zeta_R = 10^{-3}$, and convert it to an equivalent temperature $T_{\rm ERB}$ to obtain the total radio background temperature, $T_R$. This elevated background radiation leads to a deeper absorption feature, consistent with the amplitude reported by the EDGES collaboration \cite{Bowman:2018yin}.

We then explore the impact of DM annihilation heating on the global 21-cm signal. As before, we fix the annihilation cross-section at $\langle\sigma v\rangle = 10^{-26}~\rm cm^{3}s^{-1}$ and vary the dark matter mass. The blue, red, and green solid curves in Fig.~\ref{Fig:DM}b correspond to $M_\chi = 7.5$ GeV, $2$ GeV, and $0.75$ GeV, respectively. For lower DM masses, the energy injection into the IGM is enhanced, leading to a more efficient heating of the gas. This results in a suppressed absorption feature in $T_{21}$, as illustrated by the green curve. Specifically, we find that the absorption signal during the dark ages vanishes around $z \sim 89$, while at $z \sim 17$ the amplitude of $T_{21}$ is reduced to approximately $-21$ mK. Furthermore, upon closer inspection, we can find a slight leftward shift in the dark ages absorption trough, indicating an extended period of collisional coupling due to the increased gas temperature. Notably, even in scenarios where the heating from DM annihilation is sufficient to erase the dark ages signal, an absorption feature can persist during the cosmic dawn. It is worth mentioning that the analysis here excludes the contribution from an excess radio background, which, if included, would enhance the depth of the cosmic dawn absorption without affecting the dark ages signal in the presence of DM annihilation.

\begin{figure}
\centering
\includegraphics[width=\linewidth]{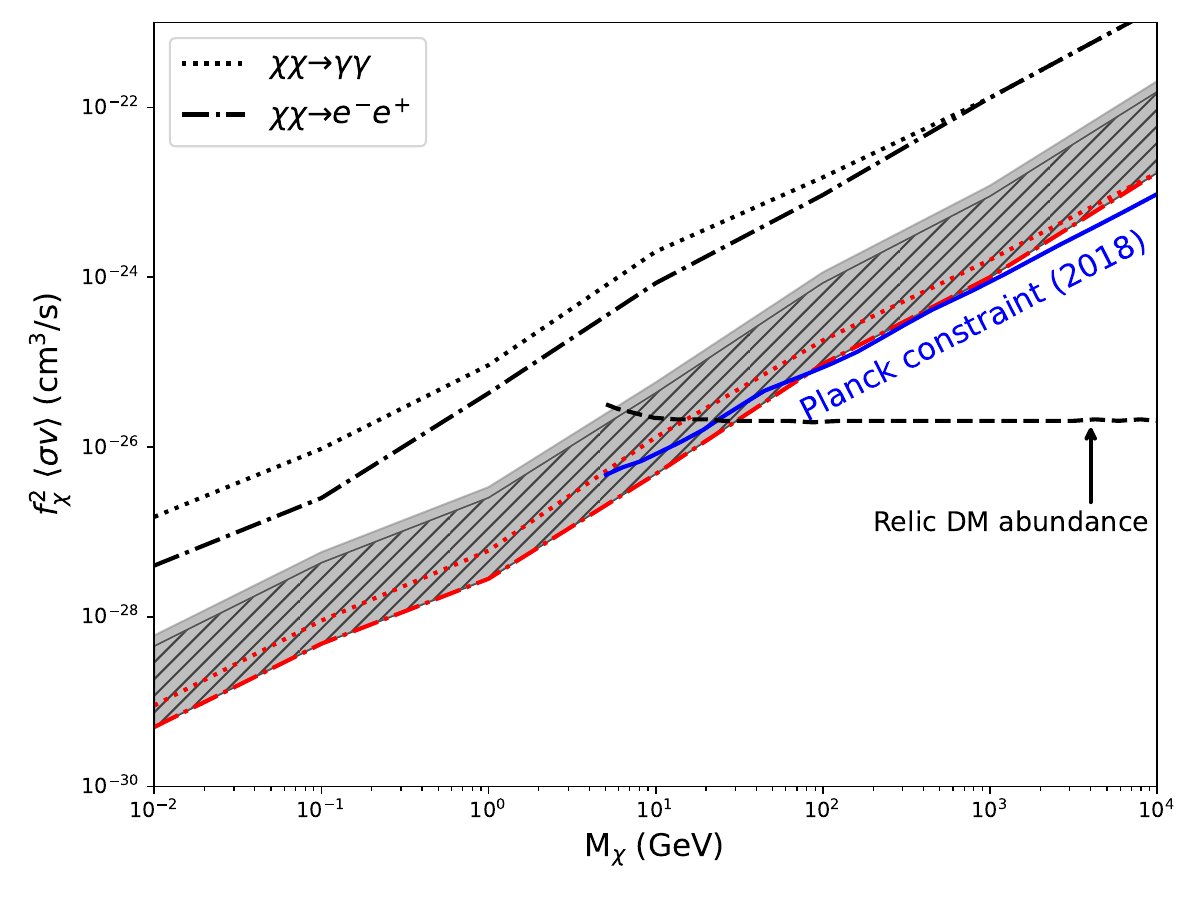}
\caption{Represent upper bounds on $f_\chi^2\langle\sigma v\rangle/M_\chi$. The black dotted and dash-dotted lines represent the upper limit on the annihilation cross-section, via $\chi\chi\to\gamma\gamma$ and $\chi\chi\to e^{-}e^{+}$ respectively, above which the dark ages signal vanishes. Whereas, the red dotted and dash-dotted lines set a lower limit of $\gtrsim 6\times 10^{-28}~\rm cm^3\,s^{-1}\,GeV^{-1}$ for $\chi\chi\to\gamma\gamma$ and $\gtrsim 2.79\times 10^{-28}~\rm cm^3\,s^{-1}\,GeV^{-1}$ for $\chi\chi\to e^{-}e^{+}$ channel, respectively, required to mildly alter the standard $T_{21}$ amplitude at $z\sim 89$. The grey shaded region depicts observational uncertainty of $\Delta\, 5\,\rm mK$ in future lunar-based experiments \cite{Burns:2020gfh}. The grey cross-hatched region shows theoretical uncertainties due to cosmological parameter variations. The blue solid line represents an upper limit from Planck (2018), which is $\lesssim 4.7\times 10^{-27}~\rm cm^3\,s^{-1}\,GeV^{-1}$ for $M_\chi>5~\rm GeV$ \cite{Planck:2018vyg}. The black dashed line presents the thermal relic cross-section of $\langle\sigma v\rangle \simeq 3\times 10^{-26}~\rm cm^3\,s^{-1}$ taken from Ref. \cite{Planck:2018vyg, Short:2019twc}.}
\label{Fig:bound}
\end{figure}

On this ground, we derive upper bounds on the DM mass and annihilation cross-section, as shown in Fig.~\ref{Fig:bound}. 
The black dotted and dash-dotted lines illustrate the upper limit on $f_\chi^2\langle\sigma v\rangle/M_\chi$ beyond which the absorption signal during the dark ages vanishes for annihilation channels $\chi\chi\to\gamma\gamma$ and $\chi\chi\to e^{-}e^{+}$, respectively. This corresponds to a constraint of $f_\chi^2\langle\sigma v\rangle/M_\chi \lesssim 9\times10^{-26}~\rm cm^3\,s^{-1}\,GeV^{-1}$ and $\lesssim 4\times10^{-26}~\rm cm^3\,s^{-1}\,GeV^{-1}$, respectively. Since the standard $\Lambda$CDM scenario predicts no luminous sources during the dark ages, any deviation from the expected $T_{21}$ absorption depth of $\sim 42\,\rm mK$ at $z \sim 89$ may indicate the presence of non-standard physics. The red dotted and dash-dotted lines indicate the lower limit of $f_\chi^2\langle\sigma v\rangle/M_\chi \gtrsim 6\times 10^{-28}~\rm cm^3\,s^{-1}\,GeV^{-1}$ for $\chi\chi\to\gamma\gamma$ and $\gtrsim 2.79\times 10^{-28}~\rm cm^3\,s^{-1}\,GeV^{-1}$ for $\chi\chi\to e^{-}e^{+}$ channels, respectively, necessary to produce a mild modification to the signal's amplitude.

We note that the electron–positron annihilation channel can provide relatively stronger bounds compared to the photon channel for dark matter masses 
$M_{\chi}\lesssim 10^3~\rm GeV$. This difference arises from the fact that the annihilation $e^{-}e^{+}$ couples more efficiently to the IGM than the $\gamma\gamma$ channel. In the $e^{-}e^{+}$ case, the injected electrons and positrons lose energy primarily through inverse Compton scattering with CMB photons, Coulomb interactions with free electrons, and collisional ionization and excitation of neutral hydrogen atoms. These processes deposit a significant fraction of the injected energy into the IGM as heat, ionization, and excitation, leading to a rise in the ionization fraction and IGM temperature. In contrast, photons produced via the $\gamma\gamma$ channel initially interact weakly with the IGM. For photon energy less than a TeV, their mean free paths are large, and they typically free-stream until redshifted or converted into secondary particles through interactions such as photon–photon pair production on the CMB or pair production on neutral hydrogen ~\cite{DAmico:2018sxd, Mitridate:2018iag, Liu:2018uzy, Liu:2019bbm, Short:2019twc}. As a result, the energy deposition efficiency $f_c(z,x_e)$ lowers significantly for $\gamma\gamma$ channel at 
$M_{\chi}\lesssim 10^3~\rm GeV$, leading to weaker effects on the IGM temperature and consequently to less stringent constraints from the global 21-cm signal. However, for $M_{\chi}> 10^3~\rm GeV$, the difference between the two channels becomes negligible. At these high energies, both photons and electrons rapidly initiate electromagnetic cascades through repeated cycles of pair production and inverse Compton scattering with CMB photons. These cascades quickly redistribute the injected energy into a spectrum of secondary electrons, positrons, and photons whose subsequent energy losses proceed in a similar process, irrespective of the initial annihilation channel. Consequently, the total energy deposition efficiency and hence the resulting heating and ionization of the IGM become nearly identical for both channels. This leads to comparable constraints on the annihilation cross-section at high masses.

As discussed in Sec.~\ref{sec:intro}, future lunar-based experiments may detect the global 21-cm signal during the dark ages with a precision of $\Delta T_{21} \sim 5~\rm mK$ (i.e., $-42^{+5}_{-5}\,\rm mK$), for an integration time of $10^5$ hours~\cite{Burns:2020gfh}. The grey shaded region in Fig.~\ref{Fig:bound} shows the parameter space for which $T_{21} \lesssim -37~\rm mK$, corresponding to a deviation beyond which the DM annihilation signatures would be observable by such experiments. Although this signal is largely free from astrophysical uncertainties, however, small changes in cosmological parameters can alter the amplitude by $\sim 6\,\rm mK$ ($-42^{+4}_{-2}\,\rm mK$), as shown in Ref.~\cite{PhysRevD.110.123506}. We conservatively adopt $T_{21} = -38~\rm mK$ as the threshold, represented by the cross-hatched grey region.
To further compare our result with the existing constraints in the literature, we present the following upper bounds. The blue solid line represents upper bounds on the $f_\chi^2\langle\sigma v\rangle-M_\chi$ parameter space by jointly analyzing Planck's CMB temperature fluctuation, polarization, gravitational lensing, and baryon acoustic oscillation data $(\rm TT,TE,EE+lowE+lensing+BAO)$ with 95\% confidence level for $\chi\chi\to \gamma\gamma$ channel \cite{Planck:2018vyg}. These constraints translate to $f_\chi^2\langle\sigma v\rangle/M_\chi \lesssim 4.7\times 10^{-27}~\rm cm^3\,s^{-1}\,GeV^{-1}$ for $M_\chi > 5\,\rm GeV$. Further, the black dashed line indicates the thermal relic cross-section of $\langle\sigma v\rangle \simeq 3\times 10^{-26}~\rm cm^3\,s^{-1}$ taken from Ref. \cite{Planck:2018vyg, Short:2019twc}. Conclusively, we can state that the dark ages signal is capable of providing stronger and astrophysical uncertainty-free upper bounds on the previously allowed parameter space, i.e., $f_\chi^2\langle\sigma v\rangle/M_\chi\lesssim 10^{-27}~\rm cm^3s^{-1}GeV^{-1}$.

\section{Conclusion and Discussion}\label{sec: conclusion}

The microscopic nature of dark matter remains an outstanding problem in the field of cosmology and particle physics. In this work, we study the effect of annihilation of dark matter particles on the evolution of dark ages global 21-cm signal. The global 21-cm signal from the dark ages is largely independent of astrophysical uncertainties, making it a robust probe for testing exotic physics in the post-recombination era. Several upcoming space- and lunar-based missions, such as \textsc{FARSIDE}~\cite{farside}, \textsc{DAPPER}~\cite{dapper}, LuSee Night~\cite{luseenight}, and \textsc{SEAMS}~\cite{seams}, have been proposed to measure this signal. For these future lunar-based experiments, an integration time of $10^5$ hours could reduce this uncertainty to $\sim 5\,\rm mK$~\cite{Burns:2020gfh, Rapetti:2019lmf}, making the detection of the standard dark ages signal feasible with high precision.

We specifically consider the annihilation channels, $\chi\chi\to \gamma\gamma$ and $\chi\chi\to e^-e^+$, and evaluate the resulting $T_{21}$ signal during the dark ages. Based on this, we present our derived upper bounds in Fig.~\ref{Fig:bound}, shown in the black dotted and dash-dotted lines for $\gamma\gamma$ and $e^-e^+$ channels, respectively, corresponding to $T_{21}\gtrsim -1~\rm mK$, whereas the red dotted and dash-dotted lines for $\gamma\gamma$ and $e^-e^+$ channels, respectively, corresponding to $T_{21} \gtrsim -42~\rm mK$. After accounting for observational uncertainties of $5~\rm mK$ (associated with future lunar-based experiments) and theoretical uncertainties of $6~\rm mK$ (due to cosmological parameter variations), we obtain a conservative upper bound of $f_\chi^2\langle\sigma v\rangle/M_\chi \lesssim 10^{-27}~\rm cm^3\,s^{-1}\,\rm GeV^{-1}$. This forecasted constraint is stronger than existing bounds on $M_{\chi}\lesssim 10~\rm GeV$, demonstrating the remarkable potential of the dark ages 21-cm signal to probe exotic energy injection scenarios. Before concluding, we note that this work adopts $\gamma\gamma$ and $e^-e^+$ annihilation channels. The bounds obtained are strong and astrophysical-uncertainty free. Additionally, exploring other annihilation channels, such as $\chi\chi\to b\bar{b}$ and $\mu^-\mu^+$, is left for future work.

\section*{Acknowledgements}
We thank Alekha C. Nayak for the discussions and valuable suggestions. We also thank the anonymous referees for their valuable suggestions and for improving the quality of the manuscript. 

\section*{Data Availability}
The primary results are generated using publicly available code \texttt{DarkHistory}~\cite{Liu:2019bbm}.

\end{document}